\documentstyle[aps,preprint]{revtex}

\def\ksi{\xi}

\def\MR{MR}
\def\MRs{MR\ }

\def\e{\epsilon}

\begin{document}
\draft

\title{Universal Spin-Induced Magnetoresistance in the
 Variable-Range Hopping Regime}
\bigskip
\author{Yigal Meir}
\address{Physics Department, Ben Gurion University, Beer Sheva 84105, ISRAEL}
\bigskip
\maketitle
\begin{abstract}
The magnetoresistance in the variable-range hopping regime due
 to Zeeman spin-splitting and intra-impurity interactions is calculated
analytically and shown to be a universal function of
$\mu H / kT \log R$. Good agreement with numerical calculations
in one and two dimensions is observed. With the inclusion of quantum
interference effects, excellent agreement with recent experiments
is obtained.
\end{abstract}
Magnetic field may have dramatic effects on the resistance of strongly
 disordered materials. In this regime, where the resistance is described
 by  Mott
law, $R= R_0 \exp\{(T_0/T)^{1\over{d+1}}\}$, with
 $T_0 \sim 1/\rho\ksi^d$,  $\rho$  the density
 of states, and $\ksi$ the localization length,
 the effects of the
 magnetic field can be divided into orbital effects and spin effects. In weak
 orbital fields, of the order of one quantum flux through the hopping area,
 $\sqrt{L^3\ksi}$, where $L$, the hopping length, is  typically
$(\ksi/T\rho)^{1/(d+1)}$, the prefactor of the exponent changes due to
 quantum interference  \nocite{nss1,nss2,sivan88,entin89,mwea,medina1,medina2}
\cite{nss1,sivan88,mwea,medina1},
  leading usually to negative
 magnetoresistance (\MR). In higher fields, of the order of one quantum flux
through
 the  area  defined by the localization length, $\ksi$ is enhanced
 \nocite{altshuler85,pichard90}\cite{altshuler85}, leading to an exponential
decrease in the resistance.
 In yet higher fields, the impurity wavefunction shrinks, resulting in an
 exponential enhancement of the resistance \cite{shklovskiibook}.

As the resistance does not depend explicitly on spin, one may not expect
 sensitivity of the resistance to spin splitting due  to magnetic field.
 However, when intra-impurity interactions are taken into account,
it was pointed out
\nocite{kamimura1,kamimura2,kamimura3,kamimura4} \cite{kamimura1}
that the polarization of
 the electron spins  will block some of the hopping processes,
 leading to an exponentially increased resistance.

It is nontrivial experimentally to separate the contributions of all
 these mechanisms to the \MR. This is easier for more disordered
 samples and at lower temperatures, where the magnetic field scales for these
 effects become more and more separated. More naturally, the technological
 progress
 in growing strongly disordered thin layers makes it possible to study only
 the spin-effects of a parallel magnetic field \nocite{faran,raikh91,
 jiang92,ohata} \cite{faran,ovadyahu95}. It is thus
 imperative to develop a good understanding of the spin effects, not only
 in order to understand transport in the presence of parallel field, but also
 in order to make it possible to separate the spin effects from the orbital
effects
 in more complicated situations. This question became even more important with
recent suggestions to use the variable-range \MRs as a sensitive magnetic
sensor
\cite{wang}.

Kamimura et al. \cite{kamimura1}, in addition to suggesting the mechanism and
 deriving the limit of high fields,
have performed a detailed numerical study of
 the dependence of the resistance on temperature and magnetic field. More
 recently, Clarke et al. \cite{clarke95}
 have proposed that the resistance may be
written,
\begin{equation}
R(H) = R_0 \exp\left\{(T_0/T)^{1\over{d+1}}F(y)\right\},
\label{R(H)}
\end{equation}
with $F$ a universal function
 of $y\equiv \mu H/T(T_0/T)^{1/d+1}$. They went on to obtain $F(y)$ in
limiting situations, and then
 calculated it numerically as a function of the dimensionless parameter $y$.

Here I report an exact calculation of the \MRs due to Zeeman
 splitting. I find that $R(H)$ is indeed given by (\ref{R(H)}), with $F(y)$
 obtained exactly in arbitrary dimensions. This function agrees very well
 with the numerical data in one and two dimensions, and with available
 experimental data \cite{ovadyahu95,jiang94}.

The starting point of this calculation is the mapping of the resistance
 problem into a percolation criterion \cite{ambegaokar} of an equivalent
 random resistor network \cite{miller}, consisting of randomly placed
 sites, of density $\rho$, with random energies $\e_i$. The resistance
 between each pair of sites is given by \cite{ambegaokar}
$R_{ij} = \exp\left\{(|\e_i|+|\e_j|+|\e_i-\e_j|)|/2T + 2 r_{ij}/\ksi
\right\}$,  where all resistances are measured in units of $R_0$.
Since the resistances vary exponentially, the overall
 resistance will be dominated by the weakest link, which is the largest
 resistance, $R$, such that
 the cluster formed by all resistances (bonds), satisfying $R>R_{ij}$,
percolates. Clearly, all states  participating in the percolating
network (occupied sites) must satisfy  $R>\exp(|\e_i|/2T)$.
 Following \cite{ambegaokar}, the percolation criterion employed here
is the following \cite{difsivan}
 --- given such an occupied site , the probability that
 a bond is attached to it has to be higher than the critical threshold, $p_c$,
 for the system to percolate,
\begin{eqnarray}
 p_c &=& {1\over {4T \log R}} \int_{-2T\log R}^{2T\log R} d\e_1\ \int d\e_2\
\rho \int  d^d r_{12}\  \Theta(R - e^{(|\e_1|+|\e_2|+|\e_1-\e_2|)/2T +
 2 r_{12}/\ksi}) \nonumber\\
&=& 3 T {{\ksi^d\rho\pi^{d/2}(\log
R)^{d+1}}\over{2^{d+1}\Gamma(d/2)d(d+1)(d+2)}} ,
\label{pc}
\end{eqnarray}
leading directly to the Mott hopping low.

In the presence of an intra-impurity interaction $U$, there are two
 types of impurities participating in the conduction process --- those whose
 energies lie in the vicinity of the Fermi energy, $E_F$ (type $A$), and those
 whose energy is around $E_F-U$ (type $B$). For $U\gg T$, the latter are
 at least singly occupied.
 Application of a magnetic field, such that $\mu H > T$, polarizes the
 singly occupied impurities, thus blocking any process of hopping from
 one singly occupied impurity to another, leading to a lower effective
 density, and to an exponential enhancement  of the resistance
\cite{kamimura1}. For
 arbitrary magnetic field one can define an equivalent resistor network,
 consisting of sites of two types $A$ and $B$, with relative densities
 $\rho_A$ and $\rho_B$, respectively, such that the resistance
 between pairs of sites is given by \cite{kamimura1,clarke95,kalia}
\begin{eqnarray}
 R^{AA}_{ij}=R^{BB}_{ij}&=&\exp\left[{{|\e_i|+|\e_j|+|\e_i-\e_j|}\over{2T}}
    + {{2 r_{ij}}\over{\ksi}} \right]\nonumber\\
 R^{AB}_{ij}=R^{BA}_{ij}&=&\exp\left[{{|\e_i|+|\e_j|+|\e_i-\e_j-2\mu
H|}\over{2T}}
    +{{\mu H}\over T}+ {{2 r_{ij}}\over{\ksi}} \right].
\label{rrn}
\end{eqnarray}
 The percolation condition (\ref{pc}) becomes
\begin{eqnarray}
\label{pch}
p_c &=& {\rho\over {4T \log R}} \int_{-2T\log R}^{2T\log R} d\e_1\ \int d\e_2\
 \int  d^d r_{12}\  \left[ p_A \rho_A\Theta(R -  R^{AA}_{ij}) +
 p_B \rho_B\Theta(R -  R^{BB}_{ij})  \right] \nonumber \\
&+& {\rho\over {4T \log R-4\mu H}}
 \int_{-2T\log R+2\mu H}^{2T\log R-2\mu H}
 d\e_1\ \int d\e_2\  \int  d^d r_{12}\  \left[  p_A \rho_B
\Theta(R -  R^{AB}_{ij}) + p_B \rho_A\Theta(R -  R^{BA}_{ij})
  \right] ,
\end{eqnarray}
where $p_A$ and $p_B$ are the fractions of the sites of types $A$ and $B$,
 respectively, on the percolating cluster. From (\ref{rrn}) and (\ref{pch})
 one finds that one can write $R(H)$ in the form (\ref{R(H)})
 with $F(x)=  \{1 /[p_a \rho_A + p_B \rho_B +
 (p_a \rho_B + p_B \rho_A)g(x)]\}^{1/d+1}$, with
 $x=2\mu H/T \log R$, and
$g(x)=\Theta(1-x)(1-x)^d[1+(d-2)x/2+(d-2)(d-1)x^2/12]/(1-x/2)$.
 An independent equation can be derived for $p_A$ and $p_B$,
\begin{equation}
{p_A\over p_B} = {\rho_A \over \rho_B} {{p_A + p_B\ g(x)}\over {p_B + p_A\
g(x)}}
\label{pab}
\end{equation}
leading to the final expression, $F(x)=\{2/[1+s(x)]\}^{1/d+1}$,
with $s(x)\equiv  \sqrt{(\rho_A-\rho_B)^2+4g(x)^2\rho_A\rho_B}$.
For small magnetic fields,
  $F(x) \simeq 1 + \rho_A\rho_Bx$, in agreement
 with perturbation theory \cite{shklovskiibook,clarke95}.  $F(x)$
 saturates at $x=1$,  or
$\mu H = T (T_0 / T \max\{\rho_A, \rho_B\})^{1/d+1}$,  at a value
 of $(1/\max\{\rho_A, \rho_B\})^{1/d+1}$, in
 agreement with Ref.\cite{clarke95}.

In Fig.1 we compare the exact result to numerical calculation of
 $\log R(H)/\log R(0)\simeq F(x)$ in one dimension. In the numerical
 calculation the resistance of a strongly disordered system of length
 $L=50\ksi$ and temperature $T=0.04 W$, where $W$ is the width of
 the energy distribution, has been calculated using the equivalent
 random resistor network. There are no free parameters in this comparison.
 In Fig. 2 we compare the exact result to
 the numerical data in two dimensions reported in
 \cite{clarke95}. Here the resistance has been calculated
 using the mapping into the percolation problem. Since the parameter $x$
 used in that work differs from ours and the ratio depends sensitively
 on the value of $T_0$, we allowed a single fitting parameter --- the
 $x$-axis scale.  In both Figs. 1 and 2,
 data has been presented for (a) $\rho_A/\rho_B = 1$,  and (b)
  $\rho_A/\rho_B = 1/2$.
 Excellent agreement between the numerical calculations and the analytic
 calculation is observed.

The real test of the theory is comparison to experimental data.
In Fig. 3 we compare the analytic result (broken curve) in 2d
to the experimental
data for the \MRs of an In$_2$O$_{3-x}$
 layer of thickness $d=110$nm in a parallel
field \cite{ovadyahu95}.
In the inset we compare the analytic result in 3d to
the experimental data extracted from \cite{jiang94}. In the latter work
the magnetic field dependence of the  the exponent (Eq.\ref{R(H)})
was attributed to an decrease in $\ksi$ (even though
theory \cite{altshuler85} predicts  that $\ksi$ increases with field). Here
 the data was replotted in terms of $F(x)$.
In both plots the two fitting parameters used were $\rho_A$ and the scale
of the x-axis. While good agreement may be observed in the two figures
at high fields,
there are clear deviations at low fields. This is due to the contribution of
 the orbital effects,  most
  importantly the quantum interference effects.

 The quantum interference effects -- the coherent
scattering of the hopping electron by all other impurities --
 have been taken
 into account  within the percolation approach in \cite{sivan88}.
The percolation condition (\ref{pc}) now also
 involves an interference probability $y^2$ multiplying the
 exponent inside the $\Theta$ function,
   and an integration over its distribution,  which can be derived
   independently\cite{sivan88,mwea}. The resulting
 integral equation for the
\MRs has to be solved numerically,  except for low
 magnetic fields or deep in the insulating regime. In the latter regime
 one finds that the resistance is multiplied by $\exp\{-<\log y^2>\}$, where
 the average is over the amplitude distribution. Using direct
 integration or random matrix arguments, $<\log y^2>$ was found to
 be \cite{mwea}
 $-\gamma-\log2+\log(1+\sqrt{2-H/H_{\phi}}H/H_{\phi})$,  where $\gamma$
 is the Euler constant,  and the
 field $H_{\phi}$ corresponds to one flux quantum through the hopping length.
  A similar procedure can be applied in the presence of spin-split states,
 as the quantum interference amplitude can be incorporated into the
 percolation condition (\ref{pch}) and into Eq.(\ref{pab}). The calculation
 does not lead to a change in $F(x)$,  but leads to a multiplicative
 factor in Eq.(\ref{R(H)}),   $\exp\{-A(x)<\log y^2>\}$,  with
$ A(x) = 1 - 4 \rho_A\rho_Bx g(x) g^{'}(x)/(d+1)s(x)(1+s(x))$.

 Interestingly,  this final result suggests that
 the prefactor resulting from the quantum interference is changing even
  for fields larger than the saturation field $H_{\phi}$,  due to the
  change in the effective number of impurities participating in
  the quantum interference.

   The resulting \MRs is plotted in Fig. 3 (solid line). The three fitting
   parameters used in this plot were $\rho_A = 0.76$,
   $(T_0/T)^{1/d+1} = 8.2$, and
   $H_{\phi} = 8.1$ Tesla (no fit was used to scale the x-axis).
    The latter two values should be compared
   to the approximate experimental values,  $(T_0/T)^{1/3} \sim 10.6$,
   and $\phi_0/dL\sim 9.1$ Tesla,  respectively.
  An almost perfect agreement with the experimental data is observed,
  especially in view of the fact that in the experimental paper
\cite{ovadyahu95}
 a 5-parameter fit was used,  and even then the agreement,  it was claimed,
  ``by itself,  was not trivial", within the physical constraints on the
parameters used. Similar procedure can be successfully applied to fit the data
 of Ref.\cite{jiang94} (inset of Fig. 3).

  For higher temperatures or perpendicular magnetic fields,  the last step
  in the calculation -- the mapping of the percolation problem into
   a log-averaging procedure,  is no longer applicable \cite{shklovskii91}.
   The formalism is still valid,  but one needs to solve the full integral
   equation discussed above.

 In this work we have reported an exact calculation of the magnetoresistance
 due to spin effects in the variable-range-hopping regime. The calculation
 compares very well with available numerical and experimental data.
 The mechanism responsible for that effect --- the blocking of hopping
 from a singly occupied impurity to another singly occupied impurity due
 to spin polarization suggests that spin-orbit scattering may play an
 intriguing role in such systems. As the temperature is lowered the
 hopping length gets larger and may eventually become larger than the
 spin-orbit scattering length. When this happens the electron may flip its
 spin upon hopping from one impurity to another, thus making it possible
 to hop from one singly  occupied impurity to another. Thus it is predicted
 that spin-orbit scattering may lead to nontrivial temperature effect and
 even to a {\sl decrease} in the resistance as the temperature is lowered.
 As such "turning-on" of spin-orbit scattering with lowering of the temperature
 has been demonstrated  for the magnetoresistance in the weakly localized
 regime \cite{millo90}, an experimental investigation of this prediction should
not
 be too difficult. We hope that this work will motivate such investigations.

{\sl Acknowledgments:} I thank Ora Entin-Wohlman,  Zvi Ovadyahu and Ned S.
Wingreen for various
discussions. I thank P. Clarke and L. Glazman for making the numerical data
of Ref. \cite{clarke95}
available,  and A. Frydman and Z. Ovadyahu for making the experimental data of
 Ref. \cite{ovadyahu95} available.


\vskip 8truecm
\leftline{\sl Figure Captions}

1. Comparison of the numerical calculation of
$\log R(H)/\log R(0)$ in one dimension
to the analytic $F(x)$, with {\sl no free parameters}.

2. Comparison of the analytic $F(x)$ to
  the numerical data extracted from Ref.\cite{clarke95} in two dimensions. The
only
  free parameter is the $x$-axis scale.

3. Comparison of the analytic result with (solid line) and without (broken
line)
including interference effects to the experimental data of \cite{ovadyahu95}
 and (in the inset) of \cite{jiang94}.
 The fitting parameters are discussed in the text.

\bibliography{yigal}
\bibliographystyle{yigal}

\end{document}